\documentstyle[11pt]{article}

\begin{document}
%  \begin{flushright} \begin{small}
%
%  \end{small} \end{flushright}
%\vspace{.5cm}n

%%%%  Title  %%%%
\begin{center}
{\large \bf  Extraterrestrial Solar Neutrino Physics} \vskip.5cm

%%%%%  Authors  %%%%
W-Y. Pauchy Hwang$^{a,}$\footnote{Correspondence Author;
 Email: wyhwang@phys.ntu.edu.tw} and Jen-Chieh Peng$^b$ \\
%%%%%  Address  %%%%
{\em $^a$Asia-Pacific Organization for Cosmology and Particle Astrophysics, \\
Institute of Astrophysics, Center for Theoretical Sciences,\\
and Department of Physics, National Taiwan University,
     Taipei 106, Taiwan\\
     $^b$Department of Physics, University of Illinois at Urbana-Champaign,\\
     Urbana, IL 61801-3080, USA}
\vskip.2cm

%%%%%  Authors  %%%%
%T. Harko\footnote{E-mail: tcharko@hkusua.hku.hk} \\
%%%%%  Address  %%%%
%{\em Department of Physics, The University of Hong Kong,
%    Pokfulam, Hong Kong}
%\vskip.2cm

%%%%%  Date  %%%%
{\small(March 22, 2010; Revised: June 22, 2011)}
\end{center}
%\maketitle

\begin{abstract}
We advocate the extraterrestrial solar neutrino physics (etSNP) as a means of
investigating solar neutrino physics (SNP). As we already know, the dominant
and subdominant (vacuum) oscillation lengths would be approximately one kilometer
and one hundred kilometers. On the other hand, we know so far that the
matter-enhanced oscillations take place only in the core of the Sun. Thus, the
etSNP, i.e. solar neutrino physics that could be extracted outside the Earth,
would assume a special unique role. The etSNP experiments include (1) a satellite
(detector) around the Earth or around the Jupiter or others (to provide the
shadow, for the matter-enhanced neutrino oscillations),
(2) during the Sun-Venus-Earth eclipse or similar, and (3) the chemical
compositions of the geology type (as in the Jupiter or in the Venus, to study
the origins of these planets). To be specific, we note that
the reactions induced by the $^8B$ solar neutrinos, in view of the sole high
energy nature ($E_\nu^{max}=14.03\,MeV$), would be most interesting in the
solar environment. Moreover, the experiments such as the chemical
compositions of the geology type (on the Venus or Jupiter) and
the matter-enhanced oscillations when the Sun-Venus-Earth eclipse, or the
Sun-Mercury-Earth eclipse, may also be interesting.

\medskip

{\parindent=0pt PACS Indices: 96.40.Tv (Neutrinos and muons);
96.30.Kf (Jupiter); 96.30.Dz (Mercury); 96.30.Ea (Venus);
95.85.Ry (Neutrino, etc.).}
\end{abstract}

%%%%%%%%%%%%%%%%%%%%%%%%%%%%%%%%%%%%%%%%%%%%%%%%%%%%%%%%%%%%%%%%%%%%%%
\section{Introduction}
%%%%%%%%%%%%%%%%%%%%%%%%%%%%%%%%%%%%%%%%%%%%%%%%%%%%%%%%%%%%%%%%%%%%%%

When we look at the eight major planets of our solar system, we
cannot stop being curious by many questions - and many puzzles to
ask. From inside out, the Mercury, the Venus, the Earth, and the
Mars, these might look like the earthlings, and then the Jupiter
and the Saturn might be mini-Suns, the other two or the so-called 
"ice" giant planets that we don't really
know. The fact that all eight planets fall in the same plane with
the majority along the same direction might indicate that they
might form in similar or related times. For those earthlings, the
Mercury, the Venus, the Earth, and the Mars, why only on Earth are
there living things?

The Sun provides the energy resources of all kind - the light, the
electromagnetic waves of different frequencies, the neutrinos, and the
cosmic-ray particles of all kinds - the main provider of the
extraterrestrial origin. Besides the light, solar neutrinos, which
come from the nuclear reactions in the core of the Sun, also carry
away a huge amount of energy. Differing from the light, solar
neutrinos, once produced, would travel up to the astronomical
distance without suffering second (weak) interactions - lost in
the vast space.

Solar neutrinos and all other neutrinos would be the thing that
would "shine" the dark world, after all the lights cease to ignite
- another life of the Universe if the Universe ceases to expand or
start to contract, according to our current knowledge of particle
physics and cosmology. Thus, it would be a lot more interesting to look at
neutrinos and antineutrinos more intimately, even though they involve
weak interactions or something weaker.

%%%%%%%%%%%%%%%%%%%%%%%%%%%%%%%%%%%%%%%%%%%%%%%%%%%%%%%%%%%%%%%%%%%%%%
\section{Solar Neutrinos}
%%%%%%%%%%%%%%%%%%%%%%%%%%%%%%%%%%%%%%%%%%%%%%%%%%%%%%%%%%%%%%%%%%%%%%

When the Sun is shining on us, a significant amount of the solar
energy get carried away by neutrinos. Solar neutrinos are elusive
because they only participate weak interactions - so almost all of
them pass away by us without being noticed. In fact, solar
neutrinos are even more elusive than antineutrinos because
charged weak interactions do not operate between solar neutrinos
and the ordinary low$-Z$ matter, i.e. break-up of light nuclei by
solar neutrinos being negligible - since these materials are made
of from the matter rather than the antimatter, but solar neutrinos
(before oscillation, or can't oscillate away) are also matter too.

Solar neutrinos come from the most important reactions in the
so-called $pp$-I chain\cite{Commins,PDG10},

\begin{equation}
p+p \to D+e^+ +\nu_e, \qquad (E_\nu^{max}=0.42\,MeV: \,
\phi_\nu=5.97\times 10^{10}cm^{-2} sec^{-1}),
\end{equation}

\begin{equation}
p+p+e^-\to D+\nu_e, \qquad (E_\nu=1.44: \,\phi_\nu=1.41\times
10^8),
\end{equation}
or from the $pp$-II chain,

\begin{equation}
^7Be+e^-\to ^7Li+\nu_e,\qquad (E_\nu=0.86\,MeV:\,
 \phi_\nu=5.07\times 10^9;\quad E_\nu=0.38:\,3.0\times 10^8),
\end{equation}
or from the $pp$-III chain,

\begin{equation}
^8B\to ^8Be^*+e^+ +\nu_e,\qquad (E_\nu^{max}=14.06;\,
\phi_\nu=5.94\times 10^6),
\end{equation}
or from the C-N-O cycle,

\begin{equation}
^{13}N \to ^{13}C+e^+ +\nu_e, \qquad (E=1.19: \, 2.88\times 10^8),
\end{equation}

\begin{equation}
^{15}O\to ^{15}N+e^+ +\nu_e, \qquad (E=1.70: \, 2.15\times 10^8).
\end{equation}
Here the neutrino fluxes $\phi_\nu$ are measured at the sea level
on Earth, in units of $cm^{-2}sec^{-1}$. Note that the $^8B$ neutrino
flux has been updated\cite{Kayser} to $(5.69\pm 0.91)\times 10^6 \,
cm^{-2}sec^{-1}$ (theoretically) or $(4.94 \pm 0.21 (stat)
^{+0.38}_{-0.34} (syst)) \times 10^6\,cm^{-2} sec^{-1}$
(experimentally). The quoted number is from PDG2010\cite{PDG10}.

Of course, the electron-like neutrinos may oscillate into muon-like
or tao-like specifies but fortunately neutral weak interactions do not
differentiate among them; other types of neutrino oscillations, so
far less likely, could be relevant though.

If we look at different planets, for example, the average distance
of the planet Jupiter from the Sun is 5.203
$a.u.$ with the Jupiter year 11.9 our years. The radius of the
Jupiter is 71,398 $km$, much bigger than the Earth's 6,378 $km$.
In terms of the mass, the Jupiter's $1.901\times 10^{27}\,Kg$ is
about 300 times than the Earth's $5.974\times 10^{24}\,Kg$. It is
believed that the composition of the Jupiter is similar to our
Sun, mostly the hydrogen plus a certain fraction of the helium.

Therefore, when solar neutrinos encounter the Jupiter, we
anticipate that the following weak interactions will dominate:
\begin{equation}
\nu + p \to \nu +p, \qquad \nu + ^4He \to \nu + ^4He,
\end{equation}
while the reaction $\nu + e^- \to \nu + e^-$ would serve as a
small correction. So, the solar neutrinos do stop at the different
planets but only with the tiny factions.

As a suggestion, we would use the so-called "elementary-particle
treatment" (EPT)\cite{EPT} to handle the neutrino-nucleus reactions such as
$\nu + ^4He \to \nu + ^4He$. In fact, under EPT, the "estimates" for most
reactions which we are talking about could be obtained easily. As a matter
of fact, our treatment for $\nu + p\to \nu +p$ is the typical EPT treatment
- the elementary-particle treatment for elementary particles themselves.

%%%%%%%%%%%%%%%%%%%%%%%%%%%%%%%%%%%%%%%%%%%%%%%%%%%%%%%%%%%%%%%
\section{Extraterrestrial Solar Neutrino Physics}
%%%%%%%%%%%%%%%%%%%%%%%%%%%%%%%%%%%%%%%%%%%%%%%%%%%%%%%%%%%%%%%

We may begin with a typical estimate on a satellite (carrying a detector of
$1\, m^3$) orbiting the Earth (to shield the Sun) to do the solar neutrino
experiments. Let's focus on the high-energy $^8B$ solar neutrino experiments
with the flux $5 \times 10^6 cm^{-2}sec^{-1}$. As seen below, typical cross
sections would be $10^{-42} \, cm^2$. The number density is estimated to be
$6.02 \times 10^{23}\, cm^{-3}$. The exposure time is adjustable and
is set to be one day. So, the event rate is
\begin{equation}
\phi_\nu \cdot \sigma \cdot "density" \cdot V\cdot T = 3 \times 10^{-6} \times 86400,
\end{equation}
with V in $m^3$ and T in day.
For example, the target would be rich in $^{37}Cl$ and after a certain days we
could check how much it produce $^{37}Ar$ (an active mode of the Davis Jr.
experiment). We could check how many nights (under the shadow of the Earth)
and how many days (as exposed to the Sun).

There are in fact many experiments similar to the above, one (dedicated)
category of extraterrestrial solar neutrino experiments, if the other experiments
cannot be "complete". The idea is that the results
under the shadow of the Earth "long enough" would be different that not
under the shadow - and with the time periods
fully adjustable (with the target volume also adjustable to some degree).

There are satellites that carry human beings or robots which can conduct
complicated experiments. In this way, the potential to do ET solar neutrino
experiments would be greatly enhanced. What we have in mind is that we eventually
have to understand the neutrino physics in some details, using solar neutrinos as
a possible avenue.

On other hand, in an experiment during the Sun-Venus-Earth eclipse, if the
change could be detectible, i.e., the volume $V$ would be that defined by
the Venus, then we get a big factor $10^{27}$. Of course, we have to think
about what could be actually detected.

In many situations, we are thinking of the extraterrestrial experiments
because of neutrino oscillations. Such physics, in view of its scale and
strength, may take several decades (of experimentation) to complete. That is
why we try to contemplate all the possibilities to map out a "complete" set
of experiments.

We may examine the scales of oscillation physics - that comes from
neutrino mixings and oscillations. We will find out that, for $MeV$ neutrinos
and the mass-squared differences $\Delta m^2$ in the range of
$10^{-5}$ to $10^{-3} eV^2$, the natural distance is about one to a hundred $km$.
For the matter-enhanced oscillation inside the Sun\cite{Kayser,PDG10} where the
electron density $N_e\approx 6\times 10^{25}/cm^3$ yields the interaction energy
of $0.75\times 10^{-5} eV^2/MeV$, an effect which may be duplicated inside the
Jupiter with considerable region (volume).

Of course, what is details in neutrino oscillations is in fact of importance.
For instance, if solar neutrinos, as born to be electron-like, oscillate
1/3 of time into muon-like or tau-like neutrinos, in the passage of the Jupiter.
Note that for solar neutrinos, no decay product of $\nu_\tau$ or of $\nu_\mu$
is accessible energetically. (Same as that sterile.) The simplest neutrino oscillation
reads\cite{Kayser}
\begin{eqnarray}
&P(\nu_\alpha\to \nu_\beta)=\delta_{\alpha\beta}-4\sum_{i>j} Re\{U_{\alpha i}^*U_{\beta i}
U_{\beta j}U_{\beta j}^*\}sin^2[1.27\Delta m_{ij}^2(L/E)] \nonumber \\
&\qquad\qquad +2 \sum_{i>j} Im\{ U_{\alpha i}^*
U_{\beta i}U_{\alpha j}U_{\beta j}^*\} sin [2.54 \Delta m_{ij}^2 (L/E)].
\end{eqnarray}
Here $\alpha$, $\beta$ flavor indices, $i$, $j$ the mass eigenstates,
$\Delta m_{ij}^2 \equiv m_i^2 - m_j^2$ is in $eV^2$, $L$ is in $km$,
and $E$ is in $GeV$.

We note that\cite{PDG10}
\begin{equation}
\mid \Delta m^2_{21}\mid \cong 7.6\times 10^{-5} eV^2;\quad
\mid \Delta m^2_{31}\mid \cong 2.4\times 10^{-3} eV^2;\quad
\mid \Delta m^2_{21}\mid /\mid \Delta m^2_{31}\mid \cong 0.032.
\end{equation}
So, for the solar neutrinos, the energy is of order $MeV$, $\Delta m^2$ would
be of order $10^{-3}eV^2$, then oscillation length would be in the order $1\, km$;
$\Delta m^2$ in $10^{-5} eV^2$, the length would be in $100\,km$. We may name them
as "dominant oscillation length (DOL)" and "subdominant oscillation length (SDL)".
These lengths justify why the "extraterrestrial solar neutrino physics (etSNP)"
has the natural place.

In fact, neutrino oscillations, plus matter-enhanced oscillations, occur all
the time. The question is how to observe them. The extraterrestrial solar
neutrino physics offers a natural way to do it.

%%%%%%%%%%%%%%%%%%%%%%%%%%%%%%%%%%%%%%%%%%%%%%%%%%%%%
\section{Some Calculations for the Jupiter}
%%%%%%%%%%%%%%%%%%%%%%%%%%%%%%%%%%%%%%%%%%%%%%%%%%%%%

We turn our attention to the interaction induced by the neutrinos, which
can be observed but the weak-interaction cross sections are generally too
small. In what follows, we do some exercise in order to set up our
"notations\cite{Hwang,EPT}".

For the neutral-current weak reaction induced by solar neutrinos
on the protons,
\begin{equation}
\nu(p_\nu)+p(p) \to \nu(p'_\nu)+p(p'),
\end{equation}
the transition amplitude is given by\cite{Hwang}
\begin{equation}
T={G\over \sqrt 2}i {\bar u}_\nu(p'_\nu)\gamma_\lambda(1+\gamma_5)
u_\nu(p_\nu) \cdot <p(p')\mid N_\lambda \mid p(p)>.
\end{equation}
We may proceed to parameterize the neutral-current matrix element
as follows\cite{Hwang}:
\begin{eqnarray}
&<p(p')\mid N_\lambda(0)\mid p(p)> \nonumber\\
=&i\bar u(p') \{\gamma_\lambda f_V^N(q^2)-{\sigma_{\lambda\eta}
q_\eta\over 2m_p}f_M^N(q^2) +\gamma_\lambda \gamma_5 f_A^N (q^2)
+{i2Mq_\lambda \gamma_5\over m_\pi^2}f_P^N(q^2) \}u(p),
\end{eqnarray}
with $q^2\equiv \vec q\,^2-q_0^2$, $q_\lambda=(p'-p)_\lambda$, and
$2M=m_p+m_n$. Here $f_V^N(q^2)$, $f_M^N(q^2)$, $f_A^N(q^2)$, and
$f_P^N(q^2)$, respectively, the (neutral-current) vector, weak
magnetism, axial, and pseudoscalar form factors. The differential
cross section is given by
\begin{eqnarray}
&{d\sigma\over d\Omega_\nu} (\nu + p \to \nu + p) \nonumber\\
=&{G^2(E'_\nu)^2\over 2\pi^2} {E'_\nu \over E_\nu} \{
[(f_V^N(q^2))^2 +(f_M^N(q^2))^2 {q^2\over 4m_p^2} +
(f_A^N(q^2))^2] cos^2 {\theta_\nu\over 2}  \nonumber\\
 & +2[(f_V^N(q^2)+ f_M^N(q^2))^2{q^2\over 4m_p^2}
 +(f_A^N(q^2))^2(1+ {q^2\over 4m_p^2})  \nonumber\\
 & +4{E'_\nu \over m_p}(1+{E_\nu\over m_p}
 sin^2{\theta_\nu\over 2})f_A^N(q^2)
 (f_V^N(q^2)+f_M^N(q^2))]sin^2{\theta_\nu \over 2} \}.
\end{eqnarray}

In the tree approximation in the standard model of particle
physics, we have
\begin{equation}
N_\lambda=(1-2sin^2\theta_W)I_\lambda^3-sin^2\theta_W Y_\lambda
+I_\lambda^{3(5)} - {1\over 2}Y_\lambda^s - {1\over
2}Y_\lambda^{s(5)},
\end{equation}
so that, for example,
\begin{equation}
f_V^N(q^2)=(1-2sin^2\theta_W)\cdot {1\over 2}(e_p(q^2)-e_n(q^2))
-sin^2\theta_W \cdot (e_p(q^2)+ e_n(q^2))-{1\over 2}f_V^S(q^2).
\end{equation}

\begin{equation}
f_M^N(q^2)=(1-2sin^2\theta_W)\cdot {1\over 2}(\mu_p(q^2)
-\mu_n(q^2))-sin^2\theta_W\cdot (\mu_p(q^2)-\mu_n(q^2)) -{1\over
2}f_M^S(q^2).
\end{equation}

\begin{equation}
f_A^N(q^2)={1\over 2}f_A(q^2) -{1\over 2}f_A^S(q^2).
\end{equation}

As a reasonable estimate, we could use $q^2\approx 0$ and neglect
all terms higher order in $q^2/m_p^2$ and $E_\nu/(2m_p)$. The
integration over $d\Omega$ yields

\begin{eqnarray}
\sigma & \cong & {G^2E_\nu^2\over \pi}\cdot \{({\bar f}_V^2+ {\bar
f}_A^2+...)(1+{2E_\nu\over m_p})^{-1}\nonumber\\
&& + (2 {\bar f}_A^2+ ...)(1+{2E_\nu\over m_p})^{-2}\}\nonumber\\
& \approx & 1.686\times 10^{-20}\cdot ({\bar f}_V^2 + 3{\bar
f}_A^2)\cdot ({E_\nu\over 1\,MeV})^2\cdot barn,
\end{eqnarray}
where ${\bar f}_V$ and ${\bar f}_A$ are suitable averages of
$f^N_V(q^2)$ and $f^N_A(q^2)$, respectively. Our formulas indicate
that $({\bar f}_V^2+3(\bar f)_A^2) \approx O(1)$.

The neutrinos could come from either the three-body modes (i.e.
the $\beta^+$ decays) or the two-body modes (such as the $\beta^+$
capture reactions). For the three-body modes, we could use the
phase factors to do very good estimates for the neutrino spectra;
we adopt this approximation in this paper.

Our estimate, from Eqs. (1)-(6), for the average flux times the
cross section, $\phi_\nu \sigma$, is given by
\begin{equation}
\phi_\nu\sigma=4.838\times 10^{-36}({\bar f}_V^2+3{\bar f}_A^2)
sec^{-1}.
\end{equation}
The average density of the Jupiter is 1.2469 $gm/cm^3$. Accordingly,
the famous product of Eq. (8) is estimated to be
\begin{equation}
\phi_\nu\sigma "n"=3.63 \times 10^{-12} ({\bar f}_V^2+3{\bar f}_A^2)
cm^{-3} sec^{-1}.
\end{equation}
The neutrino flux suitably weighted by the energy factor, measured
on the surface of the Jupiter, is
\begin{equation}
\phi_\nu=2.869\times 10^8 cm^{-2}sec^{-1}.
\end{equation}
This factor is already used before, calculated from from Eqs.
(1)-(6) adjusted by the distance from the Jupiter and the Sun.

As another estimate, we could compare how much energy the solar
neutrinos deposit in the Jupiter to that in the Earth,
\begin{equation}
({1\over 5.203})^2\times ({71,398 km \over 6,378 km})^3=51.82,
\end{equation}
modulated by small difference in the densities.

The stopping power can be calculated below:
\begin{equation}
{4\pi \over 3}R^3\cdot n \cdot \sigma \cdot \phi_\nu\cdot c^2 =
8.848 \times 10^8 Joule/sec.
\end{equation}
A large amount but distributed over the hugh volume (of the entire
Jupiter) - maybe leaving no trace at all.

One may wonder that our systems (planets) could be complicated - but in
fact not; that is why we introduce the EPT\cite{EPT} - how to visualize
the "complex" system as a simple system through the symmetries. For
example, the Jupiter, to the first approximation, would consist
hydrogen and helium (like the Sun). But in some calculations we
could approximate the system as composed of hydrogen, bound neutrons,
bound protons, and electrons - in other words, the contribution due to
the small fraction of the nuclei can be reliably estimated.

The size of the Jupiter, about a part in a thousand compared to the
Sun, means that the matter-enhanced oscillations could be visible through
the Jupiter. Since the orbit of the Jupiter is much farther than the Earth,
we imagine that the etSNP with the Jupiter could be accomplished by a satellite
surrounding the Jupiter - with the detector at the Satellite and the shade of
the Jupiter on/off at will. The experiments could be expensive but may be needed
in a "complete" of experiments in the design.

%%%%%%%%%%%%%%%%%%%%%%%%%%%%%%%%%%%%%%%
\section{The Estimates for the Venus}
%%%%%%%%%%%%%%%%%%%%%%%%%%%%%%%%%%%%%%%

To look at the Venus, the twin planet of our mother Earth, we should
and could do a lot of ET solar neutrino experiments - since it is
inside between the Sun and the Earth. The average distance of the Venus
from the Sun is 0.72333 a.u. and its radius is 6.652 $km$.

The estimate, when applied to the Venus, for the average flux times the
cross section, $\phi_\nu \sigma$, is given by
\begin{equation}
\phi_\nu\sigma=2.504\times 10^{-34}({\bar f}_V^2+3{\bar f}_A^2)
sec^{-1}.
\end{equation}
The average density of the Venus is 5.24 $gm/cm^3$. Accordingly,
the famous product of Eq. (8) is estimated to be
\begin{equation}
\phi_\nu\sigma "n"=7.8988 \times 10^{-10} ({\bar f}_V^2+3{\bar f}_A^2)
cm^{-3} sec^{-1}.
\end{equation}
The neutrino flux suitably weighted by the energy factor, measured
on the surface of the Venus, is
\begin{equation}
\phi_\nu=1.4849\times 10^{10} cm^{-2}sec^{-1}.
\end{equation}

As another estimate, we could obtain how much energy the solar
neutrinos deposit in the Venus. The estimate for the stopping power is
\begin{equation}
{4\pi \over 3}R^3\cdot n \cdot \sigma \cdot \phi_\nu\cdot c^2 =
3.464 \times 10^8 Joule/sec.
\end{equation}
Also a large amount because the distance from the Sun is much closer (than
the Jupiter). The volume of the Venus is $(6,652 \times 10^5 cm)^3$ or
$2.943 \times 10^{26} cm^3$, so each unit volume ($1\,cm^3$) would take
$8.496 \times 10^{17} sec$, a long time, to accumulate one Joule of
neutrino energy.

The importance of the etSNP using the Venus in the eclipse configuration
and thus investigating the matter-enhanced oscillations shouldn't be
underestimated. The DOL of one kilometer and the SOL of a hundred meters
means that the etSNP has of the right distance to play with - if the
matter-enhanced oscillations leave the marks through the Earth-Venus-Sun
eclipse, however small but detectible, the story would be remarkable.

%%%%%%%%%%%%%%%%%%%%%%%%%%%%%%%%%%%%%%%%%%%%%%
\section{Importance of $^8B$ Solar Neutrinos}
%%%%%%%%%%%%%%%%%%%%%%%%%%%%%%%%%%%%%%%%%%%%%%

There are several reasons why $^8B$ solar neutrinos are of special
importance. First of all, the energies of these neutrinos
($E_\nu^{max}=14.06 MeV$, see Eq. (4)) are higher than
the other neutrinos by a factor of ten. We know that the weak-reaction
cross sections in these energies are proportional to $E^2$. Secondly, with
these energies, many reactions among nuclei become energetically possible.
Thirdly, the
binding energies of the nuclei are so arranged that it matches up with
solar neutrinos with marvelous results. For example, the first few "deeply-bound"
nuclei, $^4He$ (or $\alpha$), $^{12}C$, $^{16}O$, etc. can only connect
through the neutral-current weak reaction induced by $^8B$ solar neutrinos,
$\nu + ^{12}C \to \nu + ^8Be + \alpha $ ($^8Be$ is effectively two $\alpha$)
and $\nu + ^{16}O \to \nu + ^{12}C + \alpha$. (To be elaborate in the next
section, this may be a deep statement but it turns out to be true.)

If we look up at the nuclear table with the binding energies, we would
realize immediately that these can categorize into the deeply bound nuclei,
with B.E. per nucleon more that 7 MeV, and those with the "last" nucleon
of much less binding energy. The deeply bound nuclei are also those naturally
abundant elements - with most of the abundance ratios assumed to be
constants\cite{sun}. In fact, all these change slowly with $^8B$ solar
neutrinos, in fact, only with $^8B$ solar neutrinos.

To see all these, consider\cite{Audi} for example $^{14}N$, virtually all the
weak reactions could be induced by solar neutrinos such as $\nu_e +
^{14}N \to e^- + ^{14}O $, $\nu + ^{14}N \to \nu + ^{13}C + p $,
$\nu + ^{14}N \to \nu + ^{13}N + n$, $\nu + ^{14}N \to \nu +^{12}C + d$,
$\nu +^{14}N \to \nu + ^{10}B + \alpha$, $\nu + ^{14}N \to
e^- +^{13}N + p$ ..., all by $^8B$ solar neutrinos. Thus, it is easy to
have $^{14}N$ ($A\ne 4j$) nuclei converted eventually into
$^{12}C$ ($A=4j$) nuclei, but not vice versa.

On other hand, only $\nu + ^{12}C \to \nu + ^8Be + \alpha$ by $^8B$ solar
neutrinos is allowed. So as mentioned for $^{16}O$ into $^{12}C$, or eventually
into $^4He$. (See the next section for estimates of the cross sections.) So, the net
effect is to increase the $A=4j$ nuclei, especially the $^4He$ nuclei.

At this juncture, we may introduce a new field - the "Solar Neutrino Induced Nuclear
Chemistry" (SNiNC), which would deal with the various nuclear chemistry induced by
solar neutrinos. For example, how many could $^4He$, $^{12}C$, $^{16}O$, ..., and
$^2H (d)$, $^3He$, ..., $^{14}N$, ..., and so on,
eventually stabilize under the Sun under billions of years? There are a lot of
interesting questions to ask. What is the abundance of a particular nucleus on Earth
and its relation to solar neutrinos in the long run? This sort of defines SNiNC, which
is different from "extraterrestrial" solar neutrino physics.

%%%%%%%%%%%%%%%%%%%%%%%%%%%%%%%%%%%%%%%%%%%%%%%%%%%%%%%
\section{Significance of Neutral Weak Interactions}
%%%%%%%%%%%%%%%%%%%%%%%%%%%%%%%%%%%%%%%%%%%%%%%%%%%%%%%

The various nuclei, together with their excited states, offer us the most
interesting and most complicated physical systems. Under the continuing
"shining" of solar neutrinos, the abundances of different species (nuclei)
on Earth (and other planets) are keeping changing. We should keep in mind
such facts as we go on.

The solar neutrinos, apart from $^8B$ solar neutrinos with energies
as high as 14.06 MeV, have the energies at most around 1 MeV. These low-energy
neutrinos would not induce any nuclear weak reactions for the most stable
nuclei including $p$, $^4He$, $^{12}C$, $^{16}O$, $^{20}Ne$, $^{24}Mg$,
$^{28}Si$, and $^{40}Ca$. These stable nuclei are also relevant abundant in the
solar system\cite{sun}. $p$ and $^4He$ are "absolutely stable under solar
neutrinos" while $^{12}C$, $^{16}O$, $^{20}Ne$, $^{24}Mg$, $^{28}Si$, and
$^{40}Ca$ are "$\beta-$decay stable under $^8B$ solar neutrinos"\cite{Audi}.
In other words, $\nu_e + ^{12}C \to e^- + ^{12}N $ does not have
enough energy to proceed for solar neutrinos ($E\le 14.06 MeV$); the same for
$^{16}O$, etc. only a few of them. Why not all the other nuclei?

In fact, the upper limit of 14.06 MeV means that $p$ and $^4He$ would not
disappear but would accumulate because of solar neutrinos. $^{12}C$ has one
channel $\nu + ^{12}C \to ^8Be + \alpha + \nu$ ($^8Be$ looks like two
$alpha$'s). $^{16}O$ has two channels, $\nu + ^{16}O \to ^{12}C + \alpha + \nu$
and $\nu + ^{16}O \to ^{15}N + p +\nu$. $^{20}Ne$ has three channels,
$\nu + ^{20}Ne\to ^{16}O +\alpha + \nu$, $\nu + ^{20}Ne \to ^{19}F + p +\nu$,
and $\nu +^{20}Ne \to ^{12}C + ^8Be + \nu$. Let's continue. $^{24}Mg$ has
four channels, $\nu + ^{24}Mg \to ^{20}Ne + \alpha +\nu$,
$\nu +^{24}Mg \to ^{23}Na+ p+ \nu$, $\nu + ^{24}Mg \to ^{16}O + ^8Be + \nu$,
and $\nu + ^{24}Mg \to ^{12}C +^{12}C + \nu$. etc.etc. All neutral weak interactions!!
Charge weak reactions, such as beta decays, exist but elsewhere, not here. The energy
conservation plus solar neutrino energies gives us the miracle.

These considerations should give a new beginning for the Solar-Neutrino-induced
Nuclear Chemistry (SNiNC again!!).

The cross sections can easily be calculated, because almost all of the initial
and final nuclei are spin zero and isospin zero. Using the EPT\cite{EPT}, we obtain,
for $\nu + ^{12}C \to ^8Be + \alpha +\nu$ (Energy Difference = 7.3666 MeV),
\begin{equation}
\sigma \approx {G^2 (E'_\nu)^2\over 2\pi} sin^4\theta_W \cdot\rho \approx
8.4303 \times (E'_\nu/10 MeV)^2\times 10^{-17}\cdot sin^4\theta_W\cdot \rho\cdot fm^2.
\end{equation}
Here $\rho$ some overlap integral squared and of order $O(1)$.

As a parenthetical remark, solar neutrinos never stop "shining" us; and this is
a way to induce basic nuclear change, for the better or worse. This is why one of
us speculate the primary source of cancers\cite{HwangP}. In any event, perhaps we
should look into those harmful nuclear reactions would be. Sorry there is no way
to escape from solar neutrinos.

%%%%%%%%%%%%%%%%%%%%%%%%%%%%%%%%%%%%%%%%%%%%%%
\section{The Composition of the Jupiter and of the Venus as the Geology Survey}
%%%%%%%%%%%%%%%%%%%%%%%%%%%%%%%%%%%%%%%%%%%%%%

Let's assume that the Jupiter was formed approximately at the same
time as the Sun. We also take the assumption that the Sun is the
first-generation star - to be consist primarily of the hydrogen
and the helium. In other words, Big-Bang Nucleosynthesis
(BBN)\cite{BBN} would provide the material for the Sun. Provided
that there was no major accident till the beginning of the Sun,
the chemical composition at the beginning was not far from the
BBN's:

\begin{eqnarray}
&^4He: Y_p=2(n/p)/(1+(n/p))\approx 0.25\nonumber\\
&^3He/p \approx 10^{-5} \nonumber\\
&^2D/p = (2.78\pm 0.29) \times 10^{-5} \nonumber \\
&^7Li/p =(1.7\pm 0.02 ^{+1.1}_{-0}) \times 10^{-10}.
\end{eqnarray}

To begin with, the Jupiter and the Sun would assume the same set
of values as BBN's. As time went by, the chemical composition in
the Jupiter would gradually change due to the $^8B$ solar
neutrinos (the only "high energy" solar neutrino,
$E_\nu^{max}=14.06\,MeV$ - see Eqs. (1)-(6)). In fact, the
amount of $^3He$ and $^2D$ in the Jupiter
would be depleted unless there would be some supply from outside
the Jupiter. Similar arguments could be developed for $^7Li$ with
some modifications.

Can the chemical composition of the Jupiter be measured
eventually? We think that this is an interesting question. Some day
a space mission could help to go to the Jupiter to get a sample
for experimentation. Before that, we think that the chemical
composition of the Jupiter may well be determined in a spectrum
experiment on the Earth, provided that some genius design is
involved. In other words, using BBN as a benchmark, the chemical
composition of the Jupiter would be very telling.

According to our previous discussions and the binding energies
listed below, the following reactions from $^4He$ (with a large
binding energy) are forbidden:
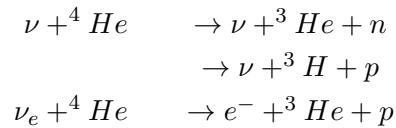
\begin{eqnarray}
\nu + ^4He & \to \nu +^3He + n   \nonumber \\
           & \to \nu +^3H  + p   \nonumber \\
\nu_e + ^4He & \to e^- +^3He + p  \nonumber \\
             & \to e^- + d + p+ p  \nonumber \\
             & \to e^- + n + p+ p+ p \nonumber \\
\nu + ^4He & \to \nu + d +p + n \nonumber \\
           & \to \nu + n+ n+ p+ p,
\end{eqnarray}
while the following reactions are possible:
\begin{eqnarray}
\nu + ^3He & \to \nu + ^2D + p \nonumber\\
           & \to \nu + n + p + p \nonumber \\
\nu_e + ^3He & \to e^- + p + p + p,
\end{eqnarray}
and

\begin{eqnarray}
\nu + ^2D & \to \nu + n + p \nonumber\\
\nu_e + ^2D & \to e^- + p + p.
\end{eqnarray}

In terms of binding energies, we have $B(^4He)=28.2956\,MeV$,
$B(^3He)=7.718 \,MeV$, and $B(^2D)=2.2245\, MeV$\cite{nucl}, thus
ruling out the possibilities for $^4He$ but keeping the reactions
on $^3He$ and $^2D$. Of course, the intermediate $n$ and $^3H$
would decay ($\beta$-decay).

Here, using the closure approximation, we obtain the cross
sections for the involved weak reactions:

\begin{eqnarray}
\sigma(\nu + A \to \nu + X) &\approx & {1\over {2\pi^2}} ({G\over
\sqrt 2})^2 4\pi<E_\nu^\prime>^2 \cdot \nonumber\\
&& \cdot \{G_V^2[(1-2sin^2\theta_W)^2+ 4sin^4\theta_W] + ... +
G_A^2 + ...\} \nonumber\\
&\approx& 1.686 \times 10^{-20}\cdot ({<E_\nu>\over 1 \,MeV})^2
\cdot 0.6041 \cdot barn.
\end{eqnarray}

\begin{eqnarray}
\sigma(\nu_e + A \to e^- + X) & \approx & {1\over {2\pi^2}}
({G\over
\sqrt 2})^2 4\pi <E_e>^2\cdot \nonumber\\
&& \cdot \{F_V^2 + ... + F_A^2 + ... \} \nonumber\\
&\approx& 1.686 \times 10^{-20} \cdot ({<E_e>\over 1\, MeV})^2
\cdot 2.6116 \cdot barn.
\end{eqnarray}

Looking into neutrino energies, the overall effects are to be
dominated by the $^8B$ neutrinos. It follows that different
numbers can then be estimated easily. We think that the scenario
reached here is very interesting indeed.

Maybe we could divide the planets into two categories: (Category I:) those
similar to the Jupiter as mini-Suns and (Category II:) those similar to the
Earth, having the elements greater than or equal to A=12. For Category I, the
discussion could stop here.

For the Venus (or planets in Category II), our other
important example, the composition is largely unknown -
maybe we could take the Earth as the profile. In the presence
of the $^8B$ neutrinos, the $^{12}C$ nucleus can change into
$^8Be$ and $\alpha$, or three $\alpha$ nuclei. This turns out
to be the most important reaction. If there are $^{16}O$ nuclei
in abundance, the $^8B$ neutrinos will change $^{16}O$ into
$^{12}C$ and $\alpha$ nuclei. If we consider these
$\alpha$-stable nuclei, the $^8B$ neutrinos provide
reactions in the anti-chain order - but slowly, more slowly
than the lifetime of the planets, but to be detectible.

At the surface of the Venus (like the Earth), there are plenty of
cosmic rays from the Sun. For example, at the surface of the Earth,
we have the intensity of nucleons from a few GeV up to above 100 TeV,
\begin{equation}
I_N(E) \approx 1.8 E^\alpha {nucleons\over cm^2\cdot sec \cdot Sr\cdot GeV},
\end{equation}
with $\alpha=\gamma +1 \approx 2.7$. Among these, 79 \% are free protons and
70 \% of the rest are nucleons bound in helium nuclei. This is another major
source of nuclear reactions which we can think of. Of course, there are some
meteorites bombarding the Earth's surface. Without some reliable estimates of
these numbers, the present paper can be safely referred to "the inside of the
Venus, and etc.".

Cosmic rays would be another sources, similar to solar neutrinos, that would
induce change in nuclear chemistry. Meteorites, astroids, and comets would be
the other. Maybe, to the first approximation, we could neglect all these.

To make the discussions easier, we may introduce two "units":
\begin{equation}
T_0=1/2\,\, billion\,\, yr=4.32 \times 10^{13} sec;
\end{equation}
\begin{equation}
\Gamma_0=1\,\, mole\times 10^{-42} cm^2 \cdot \Phi^B_\nu
= 3.011 \times 10^{-12} sec^{-1},
\end{equation}
with $\Phi^B_\nu=5\times 10^6 cm^{-2} sec^{-1}$. For example, one mole (about
1 $cm^3$) of material on the Earth would be bombarded by $^8B$ solar neutrinos
with $\Gamma_0$ interactions per second. During the Earth's life, it would be
$T_0\Gamma_0 = 130$ interactions.

These standard units indicate that the extraterrestrial solar
neutrino physics involves the reactions fairly feeble and reactional
rates fairly low. We would say that they are low-energy neutral weak
interactions - slight slow than the charged weak interactions. It is very
difficult but not impossible to achieve.

In a satellite experiment such as the dedicated Davis Jr. experiment,
we use the enriched $^{36}Cl$ as the target (on the satellite) and collect
the $^{36}Ar$ after the mission. It is clear that the experiment would be
feasible, using the above estimates as a guide.

%%%%%%%%%%%%%%%%%%%%%%%%%%%%%%%%%%%%%%%%%
\section{Matter-enhanced Neutrino Oscillations}
%%%%%%%%%%%%%%%%%%%%%%%%%%%%%%%%%%%%%%%%%

Neutrino oscillations could happen in several ways - oscillating
into different flavors but conserving the total lepton number
($L=L_e+ L_\mu+ L_\tau$), oscillating into the sterile species
($\nu_s$), oscillating into the antineutrinos via the so-called
"see-saw" mechanism, and so on. Of course, we don't know exactly
in what way neutrino oscillations take place\cite{Kayser} and for the
sake of simplicity we assume that the Nature would prefer the simplicity
and choose the first option.

Matter-enhanced neutrino oscillations is now established to be of
importance in the Sun. We don't know how big the signal when neutrinos
pass through the Venus or Mercury - to eventually measure during the
Sun-Venus-Earth eclipse or the Sun-Mercury-Earth eclipse. On the other
hand, we could speculate that what happens in the Sun is also true in the
Jupiter, the Mini-Sun, a factor of 10 smaller (in diameter).
To study the effects, we imagine that some satellite is launched to
circulate the Jupiter such that the Sun are in line with the Jupiter
and the Satellite. Similar could be thought of the Sun and the Venus and
the satellite configuration.

We think that the Sun-Jupiter-satellite experiments should be seriously
considered mainly because all our knowledge points to the positive
matter-enhanced oscillation experimental results.

The situation during the Sun-Venus-Earth eclipse would be different in detail from
that during the Sun-Mercury-Earth eclipse. Here we have used\cite{Kayser}
$\Delta m^2=8.0 \times
10^{-5} eV^2$ and $\theta=33.9^\circ$, and now use the relevant distance,
of the order $10^3km$, and the neutrino energy, of a few $MeV$; we see that the
angular factor is more than of order unity - fortunately!! It means that we can
in principle measure everything.

Here we wish to point out that it's very different from that in the so-called
"day-night" effect, done on the Earth - in terms of the phase space. The "day-night"
effect involves a small cone of the Earth and could be very small.

The question is whether the matter-enhanced neutrino oscillations could be
studied using the Sun-Venus-Earth eclipse or the Sun-Mercury-Earth eclipse.
Most of these oscillation issues might in principle be investigated
in experiments on the Earth - maybe there would be no need to go to the
Jupiter or the Venus to enhance our knowledge.

Coming to think about it, everything presumably happens in our Sun but
God forbids us from doing an experiment on the Sun except just observing. Hopefully,
to do experiments, not necessarily with solar neutrinos, on the Venus or
the Jupiter is no longer a dream, and would be a reality a few decades from now.

On the other hand, the chemical composition of the Jupiter and of
other planets, if could be measured with precision (which turns out to be very
difficult), could be another important direction to go. In particular, we may
answer the questions regarding the origins of these planets.

To sum up, the future of the extraterrestrial solar neutrino physics (etSNP) seems
to be very bright.

%%%%%%%%%%%%%%%%%%%%%%%%%%%%%%%%%%%%%%%%%%%%%%%%%%%%%%%%%%%%%%%%%%%%%%
\section*{Acknowledgments}
%%%%%%%%%%%%%%%%%%%%%%%%%%%%%%%%%%%%%%%%%%%%%%%%%%%%%%%%%%%%%%%%%%%%%%
The research of W-Y. P. Hwang is supported in part as the National Science
Council project (NSC99-2112-M-002-009-MY3), while that of J.-C. Peng is supported
by U.S. National Science Foundation.

\end{document}